\documentclass[10pt, final, conference, twocolumn]{IEEEtran/IEEEtran}
\usepackage{cite}
\usepackage{epsfig,color}
\usepackage{graphicx}
\usepackage{url}
\usepackage[centertags]{amsmath}
\usepackage{algorithm}
\usepackage{algorithmic}
\usepackage{amsfonts}
\newcommand{\beq}{\begin{equation}}
\newcommand{\eeq}{\end{equation}}
\newcommand{\beqn}{\begin{align}}
\newcommand{\eeqn}{\end{align}}

\begin{document}

\title{\huge Secure Massive MIMO Systems with Limited RF Chains}

\author{\IEEEauthorblockN{Jun~Zhu,~\IEEEmembership{Student Member,~IEEE}, Wei~Xu,~\IEEEmembership{Senior Member,~IEEE}
and~Ning~Wang,~\IEEEmembership{Member,~IEEE}}

\thanks{%
J. Zhu is with the
Department of Electrical and Computer
Engineering, the University of British
Columbia (UBC), Vancouver, Canada (e-mail: zhujun@ece.ubc.ca). W. Xu is with the National Mobile Communications Research Laboratory at Southeast University, Nanjing, China. N. Wang is with the Department of Electrical Engineering at the Zhengzhou University, Zhengzhou, China.}}
\IEEEoverridecommandlockouts

\setcounter{page}{1}

\maketitle

\begin{abstract}
In future practical deployments of massive multi-input multi-output (MIMO) systems,
the number of radio frequency (RF) chains at the base stations (BSs) may be much smaller than the number of BS antennas
to reduce the overall expenditure.
In this correspondence, we propose a novel design framework for joint data and artificial noise (AN) precoding in a multiuser massive MIMO system with limited number of RF chains, which improves the wireless security performance.
With imperfect channel state information (CSI), 
we analytically derive an achievable lower bound on the ergodic secrecy rate of any mobile terminal (MT), for both analog and hybrid precoding schemes.
The closed-form lower bound is used to determine optimal power splitting between data and AN that maximizes the secrecy rate through simple one-dimensional search.
Analytical and numerical results together reveal that
{{the proposed hybrid precoder, although suffers from reduced secrecy rate compared with theoretical full-dimensional precoder, is free of the high computational complexity of large-scale matrix inversion and null-space calculations, and largely reduces the hardware cost.}}
\end{abstract}


\section{Introduction}
The emerging massive multiple-input multiple-output (MIMO) architecture is known to support more secure transmissions, as the large-scale antenna array equipped at the transmitter (Alice) can accurately form directional and narrow information beam pointing at the intended receiver (Bob), such that the signal power level at Bob is several orders of magnitude higher than that of any incoherent passive eavesdropper (Eve) \cite{phymassive}.
Unfortunately, these benefits may vanish if Eve also employs large antenna array for eavesdropping \cite{zhu}.
In this case, even a single passive Eve can perform successful eavesdropping, if no extra efforts are made by Alice.
Since security is a critical concern for future communication systems, how to exploit the abundance of antennas offered by the large-scale array for secure massive MIMO system design is necessary and of great significance.

The first investigation of massive MIMO {{systems}} from physical layer security perspective was presented in \cite{zhu}. It was shown that a base station (BS) equipped with a large-scale antenna array is capable of generating a powerful artificial noise (AN) \cite{hmwang} covering a broad range of spatial dimensions, to mask the secret information from the passive attacker Eve. Besides, secrecy performance can be improved in multi-cell pilot contaminated scenario, even if simple {{matched-filtering (MF)}} transmission is adopted, and the number of terminals and the number of eavesdropper antennas both grow large. The work was then extended in \cite{zhu2} by considering various linear data and AN precoders, and in \cite{zhu3} by investigating the impact of hardware impairments on the secrecy performance of massive MIMO systems. 
The authors of \cite{jwang} investigated the AN-aided jamming for Rician fading channels, while the work presented in \cite{chen} compared two classic relaying schemes for secure massive MIMO transmission. The authors in \cite{zhu4} studied the power scaling law for both pilot emission and data transmission in order to achieve a secure massive MIMO transmission without the help of AN.
%

All the aforementioned {{works}} are based on the assumption that each antenna element is supported by one dedicated radio frequency (RF) chain. 
The complex baseband symbols are tuned for both amplitude and phase.
The baseband symbols are then upconverted to the carrier frequency after passing through the RF chains, whose outputs are coupled with the antennas. However, this is too costly to deploy in practice due to the use of large antenna array. {{Moreover, it hinders the operation of frequency division duplex (FDD) massive MIMO implementations, as the number of available pilot symbols is fundamentally limited by the channel coherence time \cite{liuan}.}} On the other hand, the rapid development of circuitry technology enables the high dimensional phase-only RF (or analog) processing. In \cite{analog1} and \cite{analog2}, analog precoding was employed to achieve full diversity order and near-optimal beamforming performance by iterative algorithms.
More practical constrains such as quantized phase control and finite-precision analog-to-digital (A/D) conversion are considered in \cite{analog3}. In order to further improve system performance, related literatures \cite{hybrid1,hybrid2,Liang,hybrid3} have considered hybrid schemes by combining digital and analog preocoding.
Specifically, a lower dimensional (limited by the number of RF chains) baseband precoding is employed based on the equivalent channel acquired from the product of the analog RF precoder and the channel matrix \cite{hybrid2}.
However, there lacks adequate considerations of how the limited RF chain constraint affects security, and how to design transmission strategies to improve system security under those practical constraints.

In this work, we consider secure downlink transmission of massive MIMO systems, where the BS is equipped with large antenna array and limited number of RF chains. The scenario is of great practical interest as it will be highly expensive and energy inefficient if each antenna element is connected with one dedicated RF chain. 
The main contributions of this work are summarized in the following. 1) A novel framework for linear data and AN precoding under the constraint of limited RF chains at the BS is proposed, and improved security of massive MIMO is achieved. In particular, the analog AN precoder achieves the minimum leakage, which is equivalent to the highly complex full-dimensional (pure digital) AN precoder. 
For the hybrid scheme, we show that as long as the number of RF chains is greater than the number of mobile terminals (MTs),
optimal AN precoder for leakage minimization can be obtained at a cost (complexity) much less than the full-dimensional counterpart.
2) With imperfect channel state information (CSI) acquired via uplink training, we derive an achievable lower bound on the ergodic secrecy rate of any MT for both pure analog and hybrid precoders. The bound presented in closed-form provides insights for practical secure system design, e.g., optimal power allocation \cite{PA} by one-dimensional search. 
Analytical and numerical results together reveal that {{the proposed hybrid precoder, although suffers from a security performance loss when compared with the theoretical full-dimensional precoder, avoids the high computational complexity for large-scale matrix inversion and null-space calculation, and greatly reduces the hardware cost.}}

\section{System Models}\label{s2}
\subsection{Multi-User Downlink MIMO Network Model}
  \begin{figure}
  \centering
    \includegraphics[width=3in]{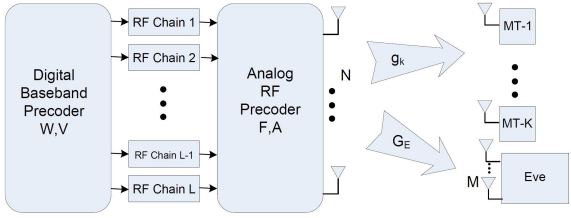}\\
    \caption{System model.}\label{sys_mod}
  \end{figure}
We consider the downlink of a multi-user MIMO system, as depicted in Fig. \ref{sys_mod}. The BS is equipped with $N$ transmit antennas driven by $L<N$ RF chains. This limits the number of transmit streams to be no more than $L$. The BS simultaneously serves $K (K < L$ and $K \ll N)$ single-antenna MTs. There is also an $M$-antenna eavesdropper, Eve, (equivalent to $M$ cooperative single-antenna eavesdroppers) located in the footprint of the BS. The goal of Eve is to extract the data for a specific MT. {{Eve works in passive mode \cite{zhu,zhu2} in order to hide her presence from the BS. \footnote{{In practice, Eve can also actively transmit jamming signals to confuse the legitimate MTs. However, this also increases her chance of being detected \cite{R1}.}}}} For downlink precoding, the BS adopts a data precoder to broadcast information-bearing signals to all MTs and an AN precoder to combat eavesdropping. Both the data and AN precoders perform digital basedband and analog RF {{processing}}. For baseband precoders ${\bf W} \in \mathbb{C}^{L_1 \times K}$ and ${\bf V}\in \mathbb{C}^{L_2 \times L_3}$, both amplitude and phase are adjustable. While RF precoders ${\bf F} \in \mathbb{C}^{N \times L_1}$ and ${\bf A} \in \mathbb{C}^{N \times L_2}$ are only phase shifters. {{The indices $L_1,L_2,$ and $L_3$ are precoder dimensions to be determined, cf. Section \ref{s4a}, and in general $L_1+L_3\leq L, L_3\leq L_2\leq L$}}.
Each entry of ${\bf F}$ and ${\bf A}$ is normalized to satisfy $|[{\bf F}]_{ij}|=|[{\bf A}]_{i,j}|=\frac{1}{\sqrt{N}}$. To this end, the received baseband signals at MT $k$ and Eve are given by
\begin{equation}
y_k={\bf g}^H_k {\bf FW} {\bf s}+{\bf g}^H_k {\bf AV}{\bf z}+n_k,
\end{equation}
and
\begin{equation}
{\bf y}_e={\bf G}^H_E {\bf FW} {\bf s}+{\bf G}^H_E {\bf AV}{\bf z}+{\bf n}_e
\end{equation}
respectively, where ${\bf g}_k =\sqrt{\beta_k} {\bf h}_k\in \mathbb{C}^{N \times 1}$ and ${\bf G}_E \in \mathbb{C}^{N \times M}$ denote the downlink channel from the BS to MT $k$ and Eve. $\beta_k$ and ${\bf h}_k$ represent the path-loss and the small-scale fading component between the BS and MT $k$. ${\bf s} \in \mathbb{C}^{K \times 1}$ with $\mathbb{E}[{\bf s} {\bf s}^H]=\frac{\phi P}{K}{\bf I}_K$ is the complex signal vector intended for the $K$ MTs, and ${\bf z} \in \mathbb{C}^{L_3 \times 1}$ is the complex AN vector with $\mathbb{E}[{\bf z} {\bf z}^H]=\frac{(1-\phi)P}{L_3}{\bf I}_{L_3}$, where $P$ denotes the total power budget and $\phi \in (0,1]$ is the power splitting factor. In order to guarantee the total power constraint, we further normalize ${\bf F},{\bf W},{\bf A},{\bf V}$ such that $\|{\bf FW}\|^2=K$ and $\|{\bf AV}\|^2=L_3$. $n_k$ and ${\bf n}_e$ are the Gaussian noise at MT $k$ and Eve, respectively. The received SINR at MT $k$ is then calculated as \cite{zhu,zhu2} $\gamma_k=$ \footnote{{The term $\mathbb{E}[|{\bf g}^H_k {\bf F} {\bf w}_k|^2]-\frac{\phi P}{K} \left(\mathbb{E}[|{\bf g}^H_k {\bf F} {\bf w}_k|]^2\right)$ at the denominator is neglected in (\ref{gammak}) as it goes to zero for $N \to \infty$ \cite[Theorem 4]{pilotcontam}.}}
\begin{equation}
\label{gammak}
\frac{\frac{\phi P}{K} \mathbb{E}[|{\bf {{g}}}^H_k {\bf F} {\bf w}_k|]^2}{\sum_{l \neq k}\frac{\phi P}{K} \mathbb{E}[|{\bf {{g}}}^H_k {\bf F} {\bf w}_l|^2]+\frac{(1-\phi) P}{L_3} \mathbb{E}[{\bf {{g}}}^H_k {\bf A} {\bf V}{\bf V}^H {\bf A}^H {\bf {{g}}}_k]+\sigma^2}
\end{equation}
for $N \to \infty$, where ${\bf w}_k (k=1,\ldots,K)$ is the $k^{\rm th}$ column of ${\bf W}$ and $\sigma^2$ is the Gaussian noise power. The corresponding achievable rate at MT $k$ is $r_k=\log_2 (1+\gamma_k)$.

\subsection{Uplink Training and CSI Acquisition}
The design of ${\bf F,W,A,V}$ requires CSI at the BS transmitters. Uplink pilot training based channel estimation is adopted due to channel reciprocity in the time-division duplex (TDD) mode, which is typical for massive MIMO systems \cite{zhu,zhu2}. Specifically, at the training phase (beginning of each coherence block), each MT transmits its own $\tau$-length pilot sequence with pilot symbol power $p_\tau$. The pilot sequences of the $K$ MTs are assumed to be orthogonal (we thus need $\tau \geq K$). Furthermore, it is reasonable to assume that the path-loss $\beta_k (1 \leq k \leq K)$ is perfectly estimated at the BS, as it changes in a much slower time scale than fast fading. When minimum mean-square error (MMSE) estimation is employed, the small-scale fading vector between the BS and MT $k$ is expressed as
\begin{equation}
{\bf h}_k=\hat{\bf h}_k+{\bf e},
\end{equation}
where the estimate $\hat{\bf h}_k$ and the estimation error ${\bf e}$ are mutually independent and can be statistically characterized as $\hat{\bf h}_k \sim \mathbb{CN}({\bf 0}_N,\lambda_k{\bf I}_N)$ and ${\bf e} \sim \mathbb{CN}({\bf 0}_N,(1-\lambda_k){\bf I}_N)$, where $\lambda_k=\frac{p_\tau \tau \beta_k}{1+p_\tau \tau \beta_{k}} \in (0,1)$. By stacking all $\hat{\bf h}_k (1 \leq k \leq K)$ together, we have $\hat{\bf H}=[\hat{\bf h}_1,\hat{\bf h}_2,\ldots,\hat{\bf h}_K] \in \mathbb{C}^{N \times K}$.

\subsection{Achievable Secrecy Rate}
A noiseless eavesdropping assumption results in an upper bound to the capacity of Eve aiming to extract the information for MT $k$, i.e., $C^k_e=\mathbb{E}[\log_2 \left(1+ \gamma^k_e\right)]$ \cite{PA}, with $\gamma^k_e=$
\begin{equation}
\label{gammake}
\frac{L_3\phi}{K (1-\phi)}{\bf w}^H_k {\bf F}^H {\bf G}_{{E}} \left({\bf G}^H_{{E}} {\bf A} {\bf V}{\bf V}^H {\bf A}^H {\bf G}_{{E}}\right)^{-1} {\bf G}^H_{{E}} {\bf Fw}_k.
\end{equation}
Based on (\ref{gammak}) and (\ref{gammake}), we obtain an achievable lower bound to the ergodic secrecy rate of MT $k$, which is $r^{\rm sec}_k=[r_k-C^k_e]^+$ \cite{zhu,zhu2}.

\section{Analog Data and AN Downlink Precoding}\label{s3}
In this section, we consider a pure analog precoder design by adopting ${\bf F} \in \mathbb{C}^{N \times K}$ and ${\bf A} \in \mathbb{C}^{N \times (L-K)}$ without ${\bf W}$ and ${\bf V}$. In this case, the achievable rate of MT $k$ reduces to $r^{\rm ana}_k=$
\begin{equation}\label{rk_ana}
\log_2 \left(1+\frac{\frac{\phi P}{K} \mathbb{E}[|{\bf {{g}}}^H_k {\bf f}_k|]^2}{\sum_{l \neq k}\frac{\phi P}{K} \mathbb{E}[|{\bf {{g}}}^H_k {\bf f}_l|^2]+\frac{(1-\phi) P}{L-K} \mathbb{E}[{\bf {{g}}}^H_k {\bf A} {\bf A}^H {\bf {{g}}}_k]+\sigma^2}\right),
\end{equation}
with ${\bf f}_l,l=1,\ldots,K$ the $l^{\rm th}$ column of ${\bf F}$.

The baseband signal (both data and AN) $[{\bf s}^H~{\bf z}^H]^H \in \mathbb{C}^{L \times 1}$ is delivered through $L$ RF chains for analog shifting before transmission, with the shifting matrix $[{\bf F}~{\bf A}] \in \mathbb{C}^{N \times L}$, where ${\bf F}$ and ${\bf A}$ will be given in Sections \ref{s3a} and \ref{s3b}.

\subsection{Analog Data Precoder Design and Performance Analysis}\label{s3a}
In this subsection, we consider the phase-only conjugate data precoder for analog data precoding. Notably, the amplitude-sensitive strategies, e.g., {{zero-forcing (ZF)}}/MMSE precoders are not considered.
For the conjugate precoder, we perform phase-only regulation at the RF domain by capturing phases of the conjugate transpose of the downlink channel from the BS to all MTs. Mathematically, we have $[{\bf F}]_{i,j}=\frac{1}{\sqrt{N}} e^{j \phi_{i,j}}$, where $\phi_{i,j}$ denotes the phase of the $(i,j)^{\rm th}$ element of the conjugate transpose of the estimated aggregate downlink channel, i.e., $[\hat{\bf h}_1,\ldots,\hat{\bf h}_K]$. The effective signal strength and the multiuser interference power in (\ref{rk_ana}) can be calculated by following \cite[Section III-B]{Liang} as
\begin{equation}\label{anasig}
\mathbb{E}[|{\bf h}^H_k {\bf f}_k|]=\frac{\sqrt{\pi N \lambda_k}}{2} \quad {\rm and} \quad \mathbb{E}[|{\bf h}^H_k {\bf f}_l|^2]=1, l \neq k.
\end{equation}
\subsection{Analog AN Precoder Design and Performance Analysis}\label{s3b}
In this subsection, we consider the phase-only iterative null-space (INS) AN precoder. Similar to Section \ref{s3}, the conventional NS AN precoder, which is commonly adopted in the literature limitations on RF chains, cf. \cite{zhu,zhu2}, is not addressed here, because the precoder is sensitive to the amplitude.
We propose a novel phase-only AN precoder by iteratively finding the NS of the aggregate MT channels. It can be verified that as long as $N$ is sufficiently large, we can always find $L-K$ (limited) sets of vectors, with each lying in the NS of the aggregate MT channels. In this case, the AN leakage power reduces to zero, with the expense of a high-computational iterative search algorithm.

\section{Hybrid Downlink Data and AN Precoding}\label{s4}
To exploit the potential of massive MIMO to the maximum extent under the limited RF chains constraint, we propose a low-dimensional multi-stream processing at the baseband by making use of the $L$ spatial degrees of freedom (DoF).
The precoded baseband signals are then upconverted to the RF chains. A phase-only control is adopted to match the $L$ RF chain outputs with the $N$ antenna elements by using phase shifters.

\subsection{Design Principles}\label{s4a}
We propose a novel hybrid data and AN precoder structure under the limited RF chain constraint, which has not been considered in the literature. It is necessary to provide some design principles and criteria beforehand, which are given as follows.

\textit{Proposition 1}: To support the data streams for $K$ MTs, we need $L_1 \geq K$. Also, to realize the NS-based AN as in \cite{zhu,zhu2}, i.e., ${\bf AV}$ is designed such that $\hat{\bf H}^H {\bf AV}={\bf 0}$, $L_2 > K$ is also required.
\begin{IEEEproof}
The former part is because the matrix $[{\bf g}_1,\ldots,{\bf g}_K]^H {\bf F}$ needs to be well-conditioned to support the multi-stream data transmission, which implies the rank of the matrix should not be smaller than the number of MTs $K$ \cite{Liang}. The latter part is due to $\hat{\bf H}^H {\bf A} \in \mathbb{C}^{K \times L_2}$.
Thus, we need $L_2 > K$ to have sufficient spatial DoF for AN transmission.
\end{IEEEproof}
Proposition 1 implies that whatever ${\bf A}$ is designed, the null-space of $\hat{\bf H}^H {\bf A}$ can always be acquired at the baseband via the digital AN precoder ${\bf V}$, without any performance loss. This is critical to simplification of the design, because ${\bf A}$ can, at least in theory, be any matrix as long as 
$\mbox{rank}[\hat{\bf H}^H {\bf A}]>K$. 

\textit{Proposition 2}: The structure ${\bf F}={\bf A} \in \mathbb{C}^{N \times L}$ with $L_1=L_2=L$ for analog phase shifter design is the best choice for achieving the highest secrecy rate.
\begin{IEEEproof}
We have $L_1=L$ because an extra data stream is always beneficial to achieve a larger data rate. On the other hand, according to \cite[Eq. (7)]{zhu2}, an extra spatial DoF adopted for AN emission is always detrimental to the eavesdropper's achieved SINR given in (\ref{gammake}), and we thus have $L_2=L$. Finally, we have ${\bf F}={\bf A}$ to simplify the design, because the choice of ${\bf A}$ does not affect the system performance, as long as the baseband AN precoder is specifically designed, according to Proposition 1.
\end{IEEEproof}

\subsection{Analog Phase Shifting}\label{s4b}
As stated in Section \ref{s4a}, the analog phase shifter can be designed in the same way as in the case with no AN. The purpose of ${\bf F}$ is thus to harvest the array gain so as to achieve multi-stream transmission. Under this criterion, we design ${\bf F}$ such that
\begin{equation}\label{Fij}
[{\bf F}]_{i,j}=\begin{cases} 1/\sqrt{N} e^{j \phi_{i,j}} & {\rm for}~1 \leq j\leq K,\\ 1/\sqrt{N} e^{j \psi}& {\rm for}~K+1\leq j\leq L,\end{cases}
\end{equation}
where $\psi$ is uniformly distributed on $[0,2\pi)$. For the limiting case of $L=K$, ${\bf F}$ in (\ref{Fij}) reduces to the form given in \cite{Liang}. For general cases, we simply generate random phases to fill the extra $L-K$ columns, which offers the extra $L-K$ spatial DoF seen from the baseband while keeping the low-complexity feature.

\subsection{Baseband Data and AN Precoder Design}\label{s4c}
For baseband processing, we perform MF and ZF data precoding and NS-based AN precoding. The precoder structure shares in common with that of the fully digital precoders, but substantially reduces the hardware complexity, because only $L$ RF chains are used, as compared to $N$ for the fully digital precoders. {{For presentation clarity, path-loss $\beta_k$ is not included in the analytical results given this section, but will exist in the resulting ergodic secrecy rate expressions in Section \ref{s5}.}}
\subsubsection{Baseband Data Precoding}
We consider two types of baseband linear data precoders in the literature, i.e., {{\cite[Section III-A]{zhu2}}}
\begin{equation}
{\bf W}=\begin{cases}\gamma_{MF} {\bf F}^H \hat{\bf H} & {\rm for~MF},\\\gamma_{ZF} {\bf F}^H \hat{\bf H} \left(\hat{\bf H}^H{\bf F}{\bf F}^H \hat{\bf H}\right)^{-1}& {\rm for~ZF},\end{cases}
\end{equation}
where $\gamma_{MF}$ and $\gamma_{ZF}$ are scaling factors satisfying the total power constraints $\|{\bf FW}\|^2=K$, which gives {{\cite[Section III-A]{zhu2}}} $\gamma_{MF}=$
\begin{equation}
\sqrt{\frac{K}{{\rm tr}\left(\hat{\bf H}^H {\bf F}{\bf F}^H {\bf F}{\bf F}^H\hat{\bf H}\right)}},\quad \gamma_{ZF}=\sqrt{\frac{K}{{\rm tr}\left(\left(\hat{\bf H}^H{\bf F}{\bf F}^H \hat{\bf H}\right)^{-1}\right)}}.
\end{equation}

For baseband MF precoding, the signal strength is calculated as 
\begin{eqnarray}\label{MFsig}
\nonumber &&\hskip-6pt\mathbb{E}[|{\bf h}^H_k {\bf F} {\bf w}_k|]\\
\nonumber &=&\hskip-6pt\gamma_{MF}\mathbb{E}[|{\bf h}^H_k {\bf F} {\bf F}^H \hat{\bf h}_k|]\\
\nonumber &=&\hskip-6pt\gamma_{MF}\mathbb{E}[|{\bf h}^H_k {\bf f}_k {\bf f}_k^H \hat{\bf h}_k|]+\gamma_{MF}\mathbb{E}[|{\bf h}^H_k \tilde{\bf F}_k \tilde{\bf F}^H_k \hat{\bf h}_k|]\\
\nonumber &\overset{(a)}{=}&\hskip-6pt \gamma_{MF}\mathbb{E}[\|\hat{\bf h}_k\| \cdot \|\hat{\bf h}_k\|]+\sum_{l \neq k} \gamma_{MF} \mathbb{E}[|{h}_{kl}|^2]\\
&\overset{(b)}{=}&\hskip-6pt \gamma_{MF} \lambda_k \left[\frac{\pi}{4}(N-1)+L\right],
\end{eqnarray}
where $\tilde{\bf F}_k \in \mathbb{C}^{N \times (L-1)}$ is by removing the column of ${\bf f}_k$ from ${\bf F}$, and ${h}_{kl}=\hat{\bf h}_k^H {\bf f}_l$. In (\ref{MFsig}), ($a$) uses (\ref{Fij}), while ($b$) exploits {{$\mathbb{E}[\|\hat{\bf h}_k\| \cdot \|\hat{\bf h}_k\|]=\frac{\pi}{4}(N-1)+1$ \cite[Section III-B]{Liang} and $\sum_{l \neq k} \gamma_{MF} \mathbb{E}[|{h}_{kl}|^2]=L-1$ due to (\ref{anasig}).}}

The scaling factor $\gamma_{MF}$ is given by
\begin{equation}
\gamma_{MF}{=}\sqrt{\frac{1}{|\hat{\bf h}^H_k {\bf F} {\bf F}^H \hat{\bf h}^H_k|}}\overset{(a)}{=}\sqrt{\frac{1}{\lambda_k\left(\frac{\pi}{4}(N-1)+L\right)}},
\end{equation}
where ($a$) uses (\ref{MFsig}). On the other hand, the intra-cell multiuser interference power is calculated as 
\begin{eqnarray}
\nonumber  &&\hskip-6pt\mathbb{E}[|{\bf h}_k^H {\bf F} {\bf w}_l|^2]\\
\nonumber &=&\hskip-6pt\gamma^2_{MF}\mathbb{E}[|{\bf h}_k^H {\bf F} {\bf F}^H\hat{\bf h}_l|^2]\\
\nonumber &=&\hskip-6pt\gamma^2_{MF} \mathbb{E}\bigg[\bigg|{\bf h}_k^H {\bf f}_k {\bf f}^H_k \hat{\bf h}_l+{\bf h}_k^H {\bf f}_l {\bf f}^H_l \hat{\bf h}_l+\sum_{j \neq k,l}{\bf h}_k^H {\bf f}_j {\bf f}^H_j \hat{\bf h}_l\bigg|^2\bigg]\\
\nonumber &\overset{(a)}{\approx}&\hskip-6pt\gamma^2_{MF} \mathbb{E}[|{\bf h}_k^H {\bf f}_k {\bf f}^H_k \hat{\bf h}_l+{\bf h}_k^H {\bf f}_l {\bf f}^H_l \hat{\bf h}_l|^2]\\
\nonumber &=&\hskip-6pt\gamma_{MF}^2 \mathbb{E}[|\|{\bf h}_k\| \cdot {h}_{lk}+\|\hat{\bf h}_l\| \cdot h_{kl}|^2]\\
&=&\hskip-6pt 2 \gamma_{MF}^2 \lambda_k \left[\frac{\pi(N-1)}{4}+1\right]
= \frac{\frac{\pi(N-1)}{2}+1}{\frac{\pi(N-1)}{4}+L} \overset{L \ll N}{\approx}2,
\end{eqnarray}
where $(a)$ uses $\sum_{j \neq k,l}{\bf h}_k^H {\bf f}_j {\bf f}^H_j\hat{\bf h}_l \ll {\bf h}_k^H {\bf f}_k {\bf f}^H_k \hat{\bf h}_l$ or ${\bf h}_k^H {\bf f}_l {\bf f}^H_l \hat{\bf h}_l$ when $K \ll N$.

For baseband ZF precoding, the signal strength is calculated as
\begin{equation}
\mathbb{E}[|{\bf h}^H_k {\bf F} {\bf w}_k|]=\gamma_{ZF},
\end{equation}
where $\gamma_{ZF}$ is given by {{\cite[Theorem 1]{Liang}}}
\begin{equation}
\gamma_{ZF}=\sqrt{\frac{K}{{\rm trace}\left(\left(\hat{\bf H}^H{\bf F}{\bf F}^H \hat{\bf H}\right)^{-1}\right)}}=\sqrt{\frac{\pi}{4}\lambda_k(N-1)}
\end{equation}
for $K \ll N$.
The multiuser interference power under ZF precoding is calculated as
\begin{equation}
\mathbb{E}[|{\bf h}_k^H {\bf F} {\bf w}_l|^2]=\mathbb{E}[|{\bf e}^H {\bf F} {\bf w}_l|^2]=1-\lambda_k,
\end{equation}
{due to $\mathbb{E}[|\hat{\bf h}_k^H {\bf F} {\bf w}_l|^2]=0$ for ZF precoder and the independence between ${\bf e}$ and ${\bf F} {\bf w}_l$.}
\subsubsection{Baseband AN Precoding}
All baseband data precoders share the same AN precoder, which is
\begin{equation}\label{avav}
{\bf V}={\rm null}\left(\hat{\bf H}^H {\bf A}\right) \in \mathbb{C}^{L \times (L-K)}.
\end{equation}
As such, the AN power leaking to MT $k$ is given by
\begin{equation}\label{AN}
\mathbb{E}[{\bf h}^H_k {\bf A} {\bf V} {\bf V}^H {\bf A}^H {\bf h}_k]={{\mathbb{E}[{\bf e}^H {\bf A} {\bf V} {\bf V}^H {\bf A}^H {\bf e}]=}}(L-K)(1-\lambda_k),
\end{equation}
{due to $\mathbb{E}[\hat{\bf h}_k^H {\bf AVV^hA^h}\hat{\bf h}_k]=0$ for the AN precoder given in (\ref{avav}), and the independence between ${\bf e}$ and ${\bf A} {\bf V}$.}

\section{Achievable Ergodic Secrecy Rate Analysis}\label{s5}
In this section, we analyze the achievable ergodic secrecy rate of the massive MIMO downlink with limited RF chains. To this end, we present an asymptotic analysis for the downlink data rate of the legitimate MTs when analog/hybrid precoding is adopted by the BS in Section \ref{s5a}. In Section \ref{s5b}, a simple closed-form upper bound for Eve's capacity is derived. The analytical bound for the resulting ergodic secrecy rate is obtained in Section \ref{s5c}. The bound is important as it has a tractable form, which allows further investigation in system design, e.g., finding the optimal power allocation that maximized the ergodic secrecy rate.

\subsection{Achievable Rate Analysis}\label{s5a}
By plugging all intermediate results obtained in Section \ref{s4c} {{including (\ref{MFsig})-(\ref{AN}) and path-loss $\beta_k$}} into (\ref{gammak}) (for hybrid precoding) and (\ref{rk_ana}) (for analog precoding), we obtain the resulting SINR achieved by MT $k$ for different types of analog/hyrbid precoding, which are summarized in Table \ref{table1}. {The effect of imperfect CSI is manifested by the parameter $\lambda_k$. When $\lambda_k=1$, all results reduce to the case of perfect CSI.}
\begin{table}
\centering
\caption{SINR of MT $k$ for analog/hybrid precoding}
\begin{tabular}{|c|c|}
  \hline
Precoder Type   & $\gamma_{k}$ \\
  \hline
  Analog (ANA) & $\frac{\frac{\phi P}{K} \left(\frac{\pi}{4} \lambda_k N\right)}{\frac{\phi P}{K} (K-1)+(1-\phi)P (1-\lambda_k)+\sigma^2/\beta_k}$\\
  \hline
  Hybrid-MF (HMF)& $\frac{\frac{\phi P}{K} \left(\frac{\pi}{4}\lambda_k(N-1)+L\right)}{\frac{\phi P}{K} 2(K-1)+(1-\phi)P (1-\lambda_k)+\sigma^2/\beta_k}$ \\
  \hline
  Hybrid-ZF (HZF) & $\frac{\frac{\phi P}{K} \frac{\pi}{4} \lambda_k (N-1)}{\frac{\phi P}{K} (K-1) (1-\lambda_k)+(1-\phi)P (1-\lambda_k)+\sigma^2/\beta_k}$\\
  \hline
  Full-MF (FMF) &  $\frac{\frac{\phi P}{K} \left(\lambda_k N\right)}{\frac{\phi P}{K} (K-1)+(1-\phi)P (1-\lambda_k)+\sigma^2/\beta_k}$\\
  \hline
  Full-ZF (FZF) & $\frac{\frac{\phi P}{K} \lambda_k (N-K)}{\frac{\phi P}{K} (K-1) (1-\lambda_k)+(1-\phi)P (1-\lambda_k)+\sigma^2/\beta_k}$\\
  \hline
\end{tabular}\label{table1}
\vskip-10pt
\end{table}
Exploiting the results in Table~\ref{table1}, we obtain the following relations between $\gamma^{\rm ANA}_k$, $\gamma^{\rm HZF}_k$, and $\gamma^{\rm FZF}_k$:
\begin{equation}
\frac{\gamma^{\rm HZF}_{k}}{\gamma^{\rm ANA}_k} \overset{N \to \infty}{\longrightarrow} 1+\frac{4{{(K-1)}}}{\pi N}\gamma^{\rm HZF}_{k},\quad \frac{\gamma^{\rm FZF}_{k}}{\gamma^{\rm HZF}_k}\overset{K \ll N}{\longrightarrow}\frac{4(N-K)}{\pi (N-1)},
\end{equation}
We note that
\begin{equation}\label{relation1}
\gamma^{\rm FZF}_k >\gamma^{\rm HZF}_k > \gamma^{\rm ANA}_k
\end{equation}
is guaranteed for $N \to \infty$ and $K \ll N$, although $\gamma^{\rm ANA}_k$ is achieved by high-computational iterative search, cf. Section \ref{s3b}.

Similarly, we obtain the relations in $\gamma^{\rm ANA}_k$, $\gamma^{\rm HMF}_k$, and $\gamma^{\rm FMF}_k$
\begin{equation}
\frac{\gamma^{\rm HMF}_{k}}{\gamma^{\rm ANA}_k}\overset{N \to \infty}{\longrightarrow}1+\frac{4(K-1)}{\pi \lambda_k N}\left(\frac{L}{K-1}- \gamma^{\rm HMF}_k\right).
\end{equation}
\begin{equation}
\frac{\gamma^{\rm HMF}_{k}}{\gamma^{\rm FMF}_k}\overset{N \to \infty}{\longrightarrow}\frac{\pi}{4}+\frac{K-1}{\lambda_k N} \left(\frac{L}{K-1}- \gamma^{\rm HMF}_k\right).
\end{equation}
Note that the achievable rate of all precoders goes large for $N \to \infty$ and $K,L \ll N$, the two conditions are unlikely to be met. To this end, we generally have
\begin{equation}\label{relation2}
\gamma^{\rm FMF}_k >\gamma^{\rm ANA}_k > \gamma^{\rm HMF}_k.
\end{equation}
\textbf{Remark 1}: By comparing (\ref{relation1}) and (\ref{relation2}), we observe that ZF precoding is desirable for hybrid design, as HZF always outperforms pure analog precoding, while HMF does not. This is mainly because HZF can fully exploit the advantage of digital design.

In particular, still according to Table~\ref{table1}, the relation between the two hybrid precoders can be obtained as
\begin{equation}
\frac{\gamma^{\rm HZF}_{k}}{\gamma^{\rm HMF}_k}=\frac{\pi \lambda_k (N-1)}{\pi \lambda_k (N-1)+4L}+\frac{4(1+\lambda_k)(K-1)}{\pi \lambda_k (N-1)+4L}\gamma^{\rm HZF}_{k},
\end{equation}
which gives $\gamma^{\rm HZF}_{k}>\frac{L}{(1+\lambda_k)(K-1)}$ to guarantee $\gamma^{\rm HZF}_k > \gamma^{\rm HMF}_k$. The discussion in this subsection will be verified in Section \ref{s6}.

\subsection{Upper Bound on Eavesdropper's Capacity}\label{s5b}
From the statistical perspective, whether analog or hybrid precoding is employed at the BS is irrelevant to the calculation of Eve's capacity, as the precoder design is independent of Eve's channel, which is unknown by the BS. Consequently, here we follow \cite[Theorem 2]{zhu} to give an identical upper bound on Eve's capacity for both the analog and hybrid precoders.
\begin{equation}\label{Cup}
C_E \leq \overline{C}_E=\log_2 \left(1+\frac{\phi M}{K(1-\phi)(1-\frac{M}{L-K})}\right)
\end{equation}
for $N \to \infty$ and $L-K>M$. We note that this bound provides an explicit relationship between Eve's capacity and various system parameters, e.g., $L,K,M,\phi$.

\subsection{Ergodic Secrecy Rate}\label{s5c}
By combining results from Sections \ref{s5a} and \ref{s5b}, we finally obtain the analytical bound for the ergodic secrecy rate achieved by any MT, which is \cite{zhu}
\begin{equation}\label{bound_rsec}
\underline{r}_k^{\rm sec}=\bigg[\log_2(1+\gamma^{\Psi}_k)-\overline{C}_E\bigg]^+,
\end{equation}
where $\gamma^{\Psi}_k$ is given in Table~\ref{table1} with $\Psi=\{{\rm ANA},{\rm HMF},{\rm HZF},{\rm FMF},{\rm FZF}\}$, and $\overline{C}_E$ is given in (\ref{Cup}).
{{Based on (\ref{bound_rsec}), we can easily find the optimal $\phi^*$ that maximizes the ergodic secrecy rate via a simple one-dimensional search, cf. Section \ref{s6},
given the ratio $\frac{M}{N}$ is smaller than a certain value that can be easily calculated for each precoding scheme by using expressions of $\gamma_k$ given in Table~\ref{table1}.}}

\section{Numerical Examples}\label{s6}
In this section, {{we verify the analytical results presented in Sections \ref{s3} and \ref{s4} through numerical and simulation results, and illustrate the impacts of the limited RF chains constraint on the system security performance. The numerical results are obtained by evaluating the analytical expression for the lower bound on the ergodic secrecy rate obtained in (\ref{bound_rsec}). Monte Carlo simulations are based on 5,000 independent channel realizations.}}
  \begin{figure}
  \centering
    \includegraphics[width=3in]{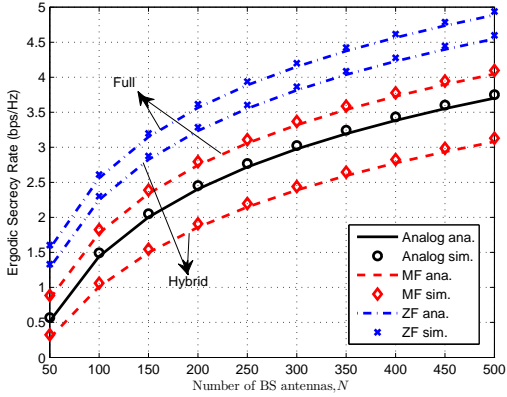}\\
    \caption{Ergodic secrecy rate comparison for different data and AN precoders, for a system with $K=3,M=5,L=10,P=10$ dB, $p_\tau=0$ dB, and $\phi=0.5$.}\label{Fig1}
  \end{figure}

In Fig.~\ref{Fig1}, we plot the ergodic secrecy rate as a function of the number of BS antennas (with fixed number of RF chains) for different pairs of data and AN precoders. We observe that the hybrid ZF precoder outperforms the pure analog precoder, which outperforms the hybrid MF precoder, when the number of BS antennas is much greater than the number of MTs and RF chains. Moreover, as expected, hybrid precoders result in a performance loss compared to their full counterparts, but have much lower computational complexity, especially for ZF. Finally, it has been verified that the lower bounds on the secrecy rate derived in Sections \ref{s4} and \ref{s5} are tight for all pairs of precoders {{by comparing the simulation results}}.
    \begin{figure}
  \centering
    \includegraphics[width=3in]{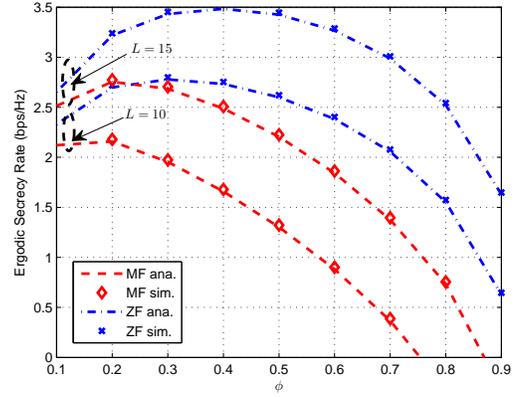}\\
    \caption{Ergodic secrecy rate comparison for hybrid precoders as functions of $\phi$, for a system with $N=128,K=3,M=5,P=10$ dB, and $p_\tau=0$ dB.}\label{Fig2}
  \end{figure}

Fig.~\ref{Fig2} compares the ergodic secrecy rate achieved by the hybrid precoders for different numbers of RF chains. The ergodic secrecy rate is plotted as a function of the power allocation factor $\phi \in (0,1)$. We observe that the secrecy rate is monotonically increasing in the number of RF chains equipped at the BS for all values of $\phi$, while the corresponding computational complexity consumed by the precoder is also expected to increase with the $L$. Another non-trivial observation is: given $L$, the optimal $\phi$ maximizing the secrecy rate for the ZF precoder is greater than that for the MF precoder. In other words, the BS inclines to allocate more resources (e.g., power) for AN transmission when MF precoder is adopted. One possible explanation is, both precoders result in identical AN leakage, cf. (\ref{AN}), while the MF precoder leads to extra multiuser interference, cf. Table \ref{table1}. Thus, it is wiser to allocate more power to data transmission when ZF precoder is used.

\section{Conclusions}\label{s7}
In this correspondence, we have proposed a novel design framework for downlink data and AN precoding in a multiuser massive MIMO system with limited number of RF chains, which improves the wireless security performance. With CSI acquired via uplink training, we derived an achievable lower bound on the ergodic secrecy rate of any MT, for both pure analog and hybrid precoders. Analytical and numerical results revealed that {{the proposed hybrid precoder, although suffers from a security performance loss when compared with the theoretical full-dimensional precoder, is free of the high computational complexity for large-scale matrix inversion and null-space calculations, and largely reduces the hardware cost.}}

\end{document}